# Choice confidence bridges credit assignment to levels of decision hierarchy

**Amir.M Mousavi Harris [1,2,8], Jamal Esmaily [3,4], Sajjad Zabbah [5,6], Reza Ebrahimpour [1,8], Bahador Bahrami [3,8]**

1. School of Cognitive Sciences, Institute for Research in Fundamental Sciences (IPM), Iran;

2. Faculty of Computer Engineering, Shahid Rajaee Teacher Training University, Iran;

3. Crowd Cognition Group Department of General Psychology and Education, Ludwig Maximillian University, Germany;

4. Graduate School of Systemic Neurosciences, Ludwig Maximilian University Munich, Germany;

5. Wellcome Centre for Human Neuroimaging, UCL Queen Square Institute of Neurology, University College London, United Kingdom;

6. Department of Computing, Goldsmiths, University of London, SE14 6NW London, United Kingdom;

7. Center for Cognitive Science, Institute for Convergence Science and Technology, Sharif University of Technology, Tehran, Iran.

8. Corresponding Author email:
amir.harris1994@gmail.com (AH)
ebrahimpour@sharif.edu (RE)
bbahrami@gmail.com (BB)

**ORCIDs:**

Amir.M Mousavi Harris: https://orcid.org/0009-0003-9847-3374

Jamal Esmaily: https://orcid.org/0000-0001-5529-6732

Sajjad Zabbah:  https://orcid.org/0000-0002-8115-3660

Reza Ebrahimpour: https://orcid.org/0000-0002-7013-8078

Bahador Bahrami: https://orcid.org/0000-0003-0802-5328

## Author Contributions

**Amir.M Mousavi Harris**: Conceptualization, Data curation, Software, Formal analysis, Validation, Visualization, Methodology, Writing – original draft

**Jamal Esmaily**: Conceptualization, Software, Validation, Visualization, Methodology

**Sajjad Zabbah**: Conceptualization, Validation, Methodology,

**Reza Ebrahimpour**: Conceptualization, Resources, Supervision, Funding acquisition, Validation, Investigation, Methodology, Project administration

**Bahador Bahrami**: Supervision, Funding acquisition, Validation, Investigation, Visualization, Writing – original draft

# Acknowledgments

BB and JE were supported by the European Research Council (ERC) under the European Union's Horizon 2020 research and innovation programme (819040 - acronym: rid-O). BB is also supported by the EMERGE EIC (Project 101070918).

# Abstract

Everyday decisions often involve many different levels. What connects these higher and lower level decisions hierarchy to one another determines how the cause(s) of failures are interpreted. It is hypothesized that decision confidence guides the assignment of blame to the correct level of hierarchy but this hypothesis has only been tested by manipulation of sensory evidence itself. We examined the consequences of modulating subjective confidence in hierarchical decision making via extra-sensory, social influence. Participants who made hierarchical, motion-plus-bandit decisions also received social information from a partner that advised the participant in the motion task. The strength of social advice - independently from sensory signals - modulated the likelihood of strategy change after negative feedback. Our findings therefore provide strong empirical evidence that subjective confidence *per se* acts as the bridge in assignment of credit and blame to various levels of decision hierarchy.

# Introduction

Buying a car is an everyday example of hierarchical decisions that we face on a day to day basis. At the lower level of this hierarchy, one has to evaluate individual features such as color, price, fuel efficiency, and safety. We also read reviews online and seek advice from friends or family members who have tried the same car model(s) and whose opinion we trust. Integrating others' experience and feedback into the decision can be very helpful. Once we have decided what car we want, there are other, higher level strategic decisions to make. Should we buy the car from Dealership A or Dealership B, based on factors such as financing terms, warranty, post-sale customer services etc. Everyday decisions often involve many different levels. In this paper we are interested in understanding what connects these higher and lower level decisions to one another? The key to this question is also to be found in our everyday experiences. If we end up, despite all due diligence and cautious deliberation, with the wrong car, how do we find out what mistake, at which level, was the cause of our trouble?

This question has been the topic of active research in behavioral and brain sciences for many decades. Early behaviorist theories such as the reflex-chain models (Loeb, 1900) suggested that a complex action like buying a car is in fact a sequence of reflexes, where one movement automatically triggers the next. Lashley (1951) (Lashley, 1951) rejected this idea and proposed that actions are organized hierarchically and controlled by central cognitive structures or programs that organize the execution of sequences in a flexible, top-down way. This top-down organization, Lashley argued, allows for corrections and adjustments when failures are encountered. But this raises an important problem. By the very nature of decision hierarchies, the causes of failures are often ambiguous.

Botvinick et al. (2009) proposed a hierarchical reinforcement learning (HRL) approach to solve this *credit assignment* problem for negative prediction error. HRL decomposes complex tasks into sub-tasks or options, so negative feedback can be assigned at the sub-task level instead of distributing blame across all actions. Attributing an error to a specific level, so that negative feedback doesn't unnecessarily affect other actions level, is key to efficient learning in HRL.

Purcell and Kiani (2016) pushed forward the idea of restricting negative feedback to the correct level of hierarchy by applying it to perceptual decision-making. They implemented a 2-level hierarchical decision paradigm consisting of a low-level direction of motion discrimination task (Britten et al., 1993) and another, higher level two-arm bandit task (Thorndike, 1898). Their hierarchical model and empirical investigations focused on sequential decision-making under uncertainty, where decisions were made across multiple levels of abstraction. They extended the concepts from HRL, such as credit assignment, to propose a mechanism for how errors and feedback are applied exclusively to the decision-making stage responsible for the error. The key factor arbitrating the credit assignment in their model was expected accuracy in the lower level, perceptual task. If a decision based on strong sensory input turned out to be incorrect, the agent would conclude that the higher level about the 2-arm bandit was wrong. Conversely, when sensory evidence was weak and expected accuracy was therefore low, then agents were less likely to switch to the other higher level bandit after negative feedback. These findings were subsequently replicated in another study involving non-human primates too (Sarafyazd & Jazayeri, 2019).

These findings raise the suggestion that metacognition may be the factor that guides the assignment of feedback. This hypothesis connects well to various other roles of metacognition. Metacognition refers to the awareness and regulation of one's own cognitive processes, and choice confidence is a key metacognitive signal that informs decision-making at different levels. Previous works have demonstrated the role of confidence in assessment of sensory vividness (Peirce & Jastrow, 1884), evidence uncertainty (Navajas et al., 2017), expected accuracy and error monitoring (Boldt & Yeung, 2015), cognitive control (Logan, 2017) and social interactions (Bahrami et al., 2010). Confidence regulates learning (Drugowitsch et al., 2019) and flexible strategic decision making in the explore-exploit dilemma (Boldt et al., 2019). These findings lend support to the hypothesis that choice confidence bridges various levels of decision hierarchy together and guides the attribution of errors to the correct level of the decision-making hierarchy.

This hypothesis makes clear and testable predictions. If a factor modulates the agent's confidence in their lower level decision in multi-level decisions such the one tested by Purcell and Kiani (2016) paradigm, then that factor should be able to change the higher level strategic decision even if it were purely extra-sensory in nature and did not affect the agent's low-level perception. The impact of such extra-sensory modulation of confidence should be qualitatively similar to changes of sensory evidence itself. For example, extra-sensory information that increases the expected accuracy should have the same impact on strategy as does increasing the sensory signal. Here we put these predictions to empirical test by modulating the agents' confidence in their perceptual judgements via social influence.

Research in perceptual and value-based decision-making has demonstrated how immediate sensory (Navajas et al., 2017) or reward-related (De Martino et al., 2013) information contributes to formation of reported confidence. This has led to the development of computational models linking confidence to signal quality, noise, and motor uncertainty (Fleming, 2024; Pouget et al., 2016). However, other more recent findings suggest a more complex view of confidence formation. In addition to immediate evidence, contextual factors such as reasoning (Mercier & Sperber, 2011), regret (Brewer et al., 2016) and social information (Bahrami et al Phil Trans 2012) can also impact confidence judgments. In the case of social information, the likelihood of making an accurate judgment is often seen as higher due to the assumption that information shared by others indicates greater likelihood of being correct. Condorcet's Jury Theorem has taught us that convergent independent opinions reduce uncertainty about the underlying phenomenon, increasing confidence (De Condorcet, 2014). In addition to its impact on accuracy, social consensus offers extra benefits (El Zein et al., 2019). Shared responsibility can increase confidence by reducing regret and enhancing our ability to justify our choices (Harvey & Fischer, 1997; Nicolle et al., 2011).

A clear example of the extra-sensory impact of social information on subjective confidence in a perceptual judgment was demonstrated by (Pescetelli et al., 2016). They studied the relative contributions of stimulus-specific and social-contingent information on confidence formation. Dyads of participants made threshold-level visual perceptual decisions, first individually and then together by sharing their wagers in their decisions. Through independent manipulation of sensory evidence to the interacting observers, they found that decision confidence was additively and linearly affected by both sensory and social information. Moreover, they observed that participants' agreements and disagreement with the partner affected their confidence in opposite directions. Given these features, we hypothesized, social information could act as the extra-sensory information that when given to the observer, could modulate their low-level choice confidence and thereby modify higher level strategy.

Therefore, to test the role of confidence in attribution of negative feedback in hierarchical decisions, we combined (Pescetelli et al., 2016) approach with Purcell and Kiani's hierarchical decision making paradigm (Purcell & Kiani, 2016). Participants made hierarchical decisions by combining sensory visual input, social information from an advising partner and trial by trial feedback.

To anticipate our results, we replicated Purcell and Kiani's (Purcell & Kiani, 2016) original finding that following a negative feedback, changing of decision strategy in the bandit task depended on the level of sensory evidence. We then replicated Pescetelli et al's (Pescetelli et al., 2016) findings that both sensory evidence and social advice modulated the participant's confidence in their lower level, perceptual decision. Critically, putting the two streams of research together, we found that social advice - independently from sensory signals - modulated the confidence in perceptual choice which, in turn, determined the likelihood of strategy change after negative feedback. Our findings therefore provide strong evidence to support the hypothesis that subjective confidence acts as a bridge in assignment of credit and blame to various levels of decision hierarchy.

## Methods

In the interest of brevity, here we describe the key elements of methods. Details are found in the supplementary material.

Following previous works (Purcell & Kiani, 2016; Sarafyazd & Jazayeri, 2019), we treated each participant as a separate experimental replication of the key hypotheses. This is because the testing of our hypothesis rests on having an adequate number of error trials under low as well as high coherence conditions. Errors under high coherence are rare. Consequently, the standard convention in human behavioral studies, i.e., collecting a couple of hundred trials from several 10s of participants would not produce the necessary statistical power. Four participants, two men and two women (average±std: 24±5 years) with normal or corrected to normal vision were recruited.

The study consisted of 3 main phases. Each participant started with a perceptual, visual motion discrimination task in which they judged the direction of motion of a random dot stimulus (Figure 1A) and expressed their confidence in their judgment. After the brief motion stimulus (500ms), two horizontal rectangles were displayed, one on each side of the center. They represented the leftward and rightward choice options. To report the direction of motion and their confidence simultaneously, the participant used the mouse to click inside one of the two colorful bars. Six confidence levels were embedded and color coded into each bar as displayed in the bottom panel of Figure 1B. Each participant took at least 1000 trials of training and continued until they achieved criterion performance level.

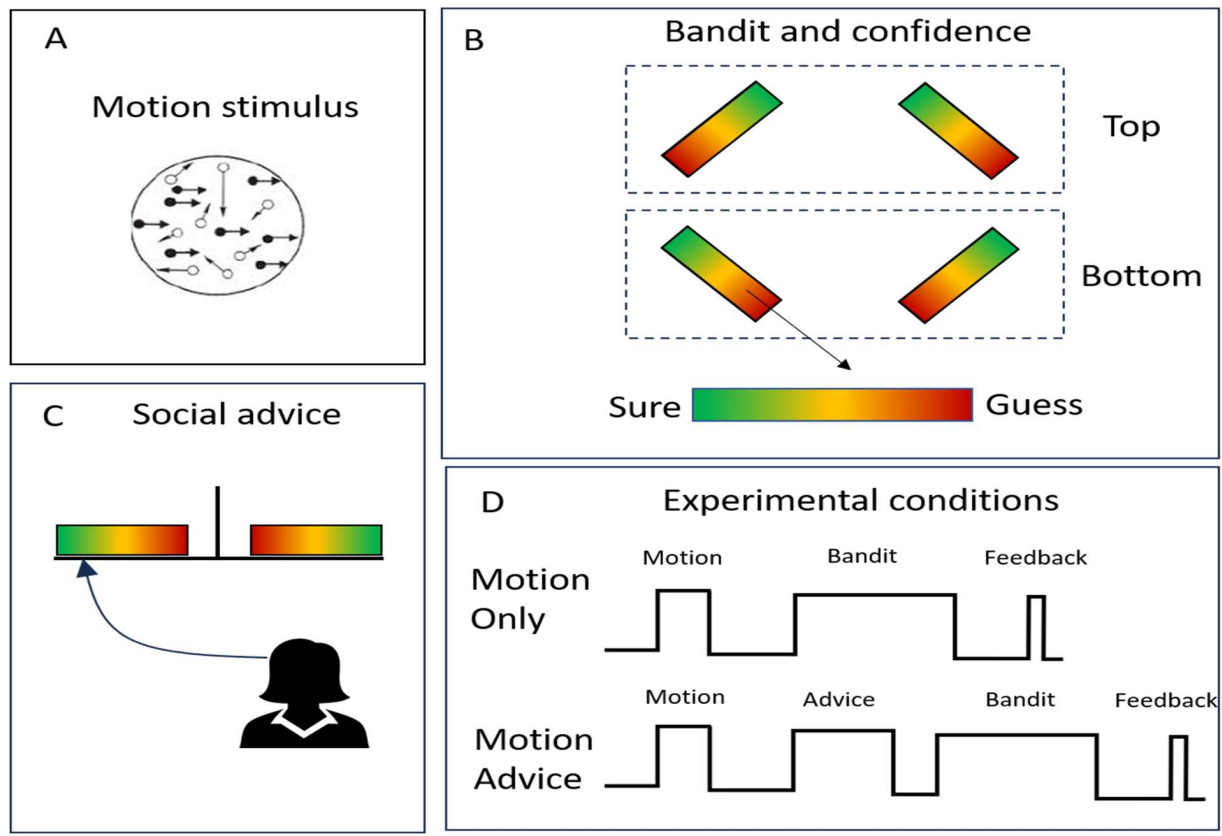

**Figure 1. Experimental paradigm**. (A) Motion stimulus. The random dot motion stimulus consisted of a field of randomly positioned dots that moved in various directions. The proportion of dots moving in the same direction varied from trial to trial and participants decided if the overall direction of motion was leftward or rightward. (B) Bandit and confidence. Within each environment, (top and bottom dashed rectangles), two sets of confidence color

bars were displayed, one for each motion direction. During training, participants had had extensive practice with confidence bars (without bandits) to familiarize with the conventions. (C) Social advice. The partner's choice of direction and level of confidence were presented graphically to the participant. (D) Sequence of events in the two experimental conditions. In the Motion-Only condition, motion stimulus was followed by the bandit and confidence display. After choice, participants received auditory feedback. In the Motion-plus-Advice condition, motion stimulus was followed by social advice, the bandit and confidence display, and finally feedback.

Having completed the training, the participant proceeded to the "Motion Only" condition (Figure 1D, top). Here, the participant had to combine their decision in the motion discrimination task with another choice in a 2-arm bandit task (Figure 1B). This 2-arm bandit was called "The Environment". At the end of the motion stimulus, the participant was given two environments (i.e., two sets of rectangles) one above and the other below the display center. Within each environment, it was possible to report the direction of motion and confidence in exactly the same manner that the participant had learned in training only here the rectangles were placed at oblique angles. The participant task was therefore two fold: to report their perceptual choice and confidence in the correctly chosen environment. The environment choice was essentially a 2-arm bandit reversal learning task. Every few trials, the correct environment changed spontaneously. Correct feedback would indicate that both motion and bandit choices were correct. If either (or both) choice were incorrect, however, feedback would be negative. Negative feedback, therefore, was ambiguous. Each participant completed 1200 trials.

Finally, participants proceeded to the Motion-plus-Advice condition (Figure 1C-D). Here, each participant was paired with four (alleged) anonymous partners. Participants were told that advice came from another participant who was simultaneously taking part in the experiment. However, they were, in fact, paired with an algorithm that produced probabilistic choice and confidence responses drawn from distributions predetermined by the experimenter. The participants were not aware of this arrangement. The overall setup and stimulus presentation remained unchanged from the Motion Only task. The key change was that after the motion stimulus, the partner's judgment of motion direction and confidence (Figure 1C) was presented to the participant for 1000ms. After advice, choice options were displayed within the environments, allowing participants to provide their own responses similar to the Motion Only condition. The participant's choice of direction of motion would, therefore, either confirm or reject the social advice. Each participant also contributed 1200 trials in this phase. Throughout all phases of the experiment, eye tracking was used to ensure that participants maintained fixation during the motion stimulus presentation.

To test our hypotheses, we employed several regression models (fitting by maximum likelihood method), whose specific details can be found in Tables S1-S7 of the supplementary materials. This approach allowed us to disentangle the effects of various factors and examine the independent contributions of each predictor separately.

# Results

The data from the Motion Only condition replicated the key behavioral findings from Purcell and Kiani (2016). Most importantly, switching between environments occurred following negative feedback and the likelihood of switching was higher when motion coherence had been higher (Fig. 2A and Table S3; Eq. 2, $\beta_1 = 4.6 \pm 0.783$, $P < 0.001$), which corresponded to trials where expected accuracy and confidence were higher (Fig. S1.A. and Table S1; $\beta_1 = 17.73 \pm 0.826$, $P < 0.001$). In addition, in supplementary material we report replicating the impact of consecutive errors on switch probability, the probability of environment switches following negative feedback since the last correct switch, and the resetting of switch evidence after positive feedback. These findings reaffirm the importance of the three key factors for strategy change: negative feedback, choice confidence, and the number of consecutive errors.

The data from the Motion-plus-Advice condition replicated Pescetelli et al.'s (2016) findings that both sensory evidence (coherence) and social advice (confirmation, rejection) modulates the participants' confidence in their low-level perceptual decisions (Fig. 2.B. and Table S6; $\beta_5 = 0.125 \pm 0.012$, $P < 0.001$; Fig. S1.F. and Table S1; $\beta_1 = 5.2948 \pm 1.073$, $P < 0.001$). We also observed an increase in both the accuracy and confidence compared to the Motion-Only task (Fig. S1.A. and Table S1; $\beta_3 = 0.249 \pm 0.074$, $P < 0.001$; Fig. S1.B. and Table S2; $\beta_2 = 0.0305 \pm 0.006$, $P < 0.001$). These findings demonstrate that social advice modulated the perceptual decision and confidence.

Having independently replicated the two foundational previous works, we then proceeded to test our key hypothesis. A logistic regression analysis confirmed that alignment with social advice increased the probability of switching after errors: conditioned on negative feedback, choices that confirmed (vs rejected) the social advice were more likely to be followed by switching between environments (Fig. 2C and Table S7; $\beta_6 = 1.38 \pm 0.19$, $P < 0.001$). Comparison of the results between the two conditions (light gray circles in Figure 2C) suggested that confirmatory and rejected social information contributed separately to strategic decisions. On one hand, having confirmed the advice (vs relying only on sensory motion information) increased the probability of switching after. On the other hand, having rejected advice and faced an error, participants were less likely to switch their strategy. This indicates that they probably interpreted the contradictory social advice as evidence that the error was unlikely to have been due to change of environment.

Finally, we found that the strength of confirmed social advice, independent of sensory signals, modulated the likelihood of a strategy change following negative feedback (Figure 2D). Advice confidence significantly influenced subjects' confidence only when the participant had chosen to confirm the advice (Table S9; $\beta_2 = 0.058 \pm 0.006$, $P < 0.001$). Importantly, rejected advice confidence did not affect the likelihood of switching (Table S8; $\beta_2 = 0.006 \pm 0.008$, $P = 0.443$). This finding is important because it is consistent with the hypothesis that social information does not *directly* influence the probability of switching but instead, it only affects the participant's confidence in their low level decision.

Consistent with the idea that social advice directly affects the probability of switching, we found that the partner's accuracy, unlike the partner's confidence, had a significant impact on the probability of switching (Table S7; $\beta_3 = 0.37 \pm 0.16$, $P < 0.05$). Interestingly, it did not influence confidence (Table S6; $\beta_3 = 0.006 \pm 0.008$, $P = 0.409$).

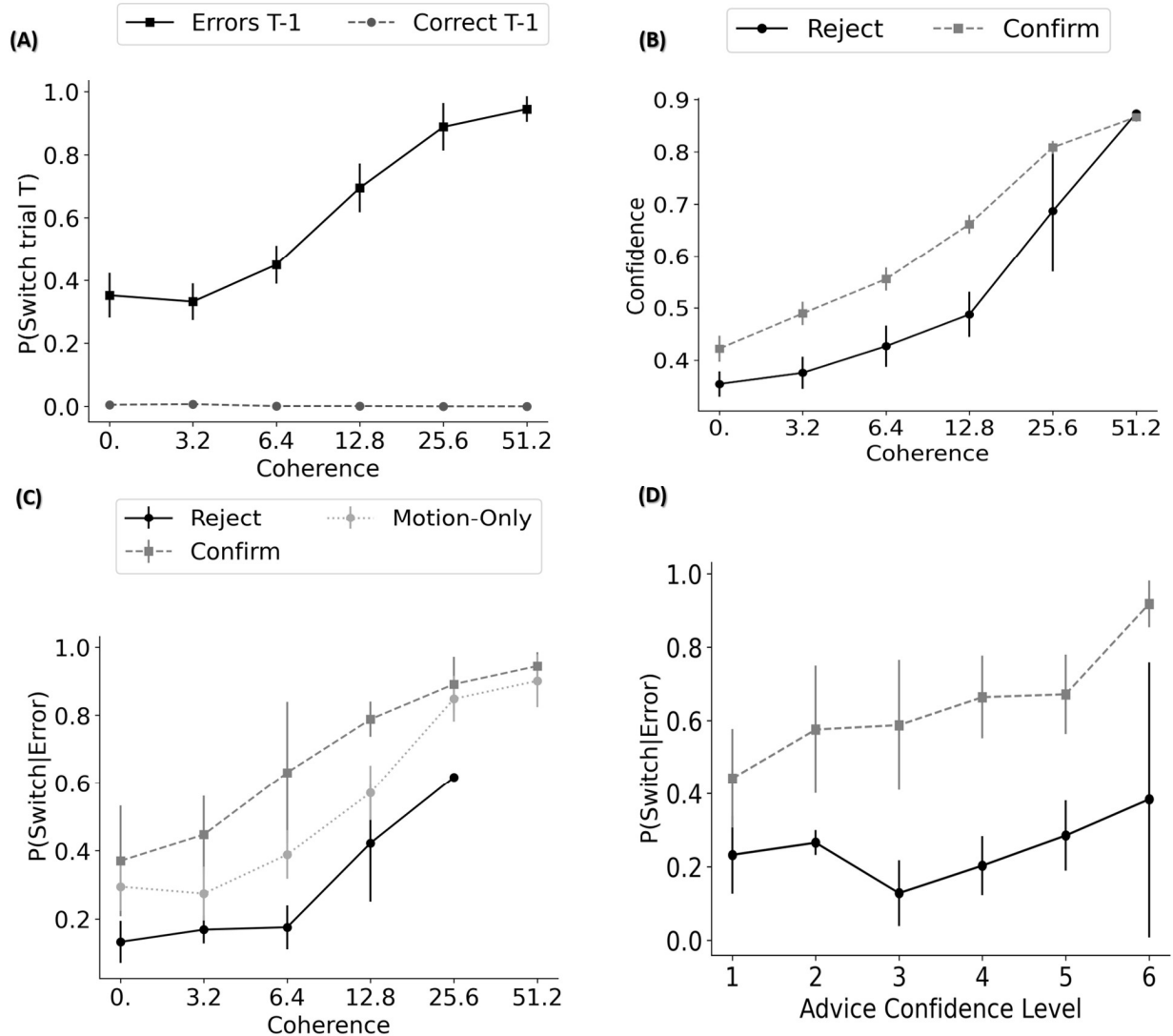

**Figure 2. Behavioral results.** (A) Only after negative feedback, the probability of switching between environments increases with the motion coherence level. (B) Social and perceptual evidence impacted the participants' confidence. Subjects' confidence levels increase with coherence and are higher when the participant confirmed (vs rejected) the partner's advice. (C) The combined effect of coherence and advice on the probability of switching after error. For better comparison, Motion-Only data from panel A is reproduced here in light gray. Note that the GLM partialed out the influence of subjects' confidence and coherence levels. (D) Switching probability increases with confirmed (but not rejected) advice confidence. Error bars are Standard Error of Mean (SEM) across subjects. Panels B-D share the same legend.

## Discussion

Human ability to undertake complex, multi-level decisions requires sophisticated hierarchical reasoning processes (Balaguer et al., 2016; Huys et al., 2015; Lake et al., 2017; Ramadan et al., 2024; Tenenbaum et al., 2011; Zylberberg, 2021). Hierarchical reasoning combines information from different levels of processing to arrive at any given choice. The outcome of this choice is, therefore, a consequence of many ingredients arising from different levels of the cognitive hierarchy. This raises the problem of credit assignment: upon encountering a negative feedback, how does the system decide which level of hierarchy to update? Here, we provided evidence supporting the theory that subjective confidence guides the assignment of errors to the appropriate level. We put forward and empirically tested the prediction that manipulating the agent's confidence in their lower-level decision should affect the higher-level strategic decision even when the manipulation is purely extra-sensory and does not alter the content of low level perception. The effect of this extra-sensory modulation of confidence on strategy selection should, we predicted, mirror the influence of sensory evidence itself.

We empirically tested this prediction by manipulating participants' confidence in their perceptual judgments through social influence. In daily life, seeking or accepting advice from others is common, and social input can significantly impact subjective confidence, potentially leading to profound changes in final decisions. To this end, we combined a social perceptual decision making paradigm (Bahrami et al., 2010; Pescetelli et al., 2016) with another, hierarchical decision-making paradigm (Purcell & Kiani, 2016). In our main experimental condition, participants undertook a multi-level, hierarchical decision task in which they received sensory evidence and social advice about the low level component of the multi-level decision.

We replicated the original findings of both studies that had inspired our question. Replicating (Purcell & Kiani, 2016) we showed that changes in decision strategies after negative feedback were influenced by the strength of sensory evidence. We also corroborated the results of (Pescetelli et al., 2016), demonstrating that sensory evidence and social advice additively modulated the participants' confidence in their low-level perceptual decisions.

Having reliably replicated both previous sets of findings, we then tested our key hypothesis by integrating the two streams of research. We showed that social advice, independent of sensory signals, influenced the participants' confidence in the low-level perceptual choices, and impacted the likelihood of strategy changes following negative feedback. Importantly, when we collapsed across all other factors, advice confidence *per se* did not modulate hierarchical decisions. If advice confidence was independently and directly incorporated into hierarchical decisions, then we would expect to see that, as advice confidence goes up, the probability of switching should also go up. But this was not the case. What we found is that advice's confidence interacts with the participant's decision to confirm or reject the advice (Figure 2C). For example, when sensory evidence was weak and the participant had rejected the social advice, probability of switching after negative feedback was substantially reduced compared to non-social baseline condition. This, we suggest, indicated that the social advice had reduced the participant's confidence in their own perceptual decision. When negative feedback followed such a situation, the participant was more likely to attribute the blame to their own erroneous low-level perception and stick to the same high level strategy. These findings provide compelling evidence that the participant's subjective confidence - and not just the strength of sensory evidence - serves as a critical mediator in assigning credit or blame within different levels of the decision hierarchy.

Confidence reflects the subjective belief, prior to receiving feedback, that a decision is correct (Fleming & Lau, 2014; Henmon, 1911; Kiani et al., 2014; Peirce & Jastrow, 1884). The alignment between this expectation and the actual feedback serves as a learning mechanism (Drugowitsch et al., 2019; Vickers, 2014), informing decisions about strategy updates. Confidence is critically involved in optimal cue integration (Deroy et al., 2016; Fetsch et al., 2012), arbitration between competing systems for behavioral control (Daw et al., 2005), guiding sequential decisions in the absence of immediate feedback (Kiani & Shadlen, 2009) and optimal interpersonal decision making (Bahrami et al., 2010; Bang et al., 2017; Esmaily et al., 2023). Our work builds up on these previous findings and goes beyond the recent reports that demonstrated the role of expected accuracy – driven by directly manipulating the low level sensory evidence – in hierarchical decision making (Purcell & Kiani, 2016; Sarafyazd & Jazayeri, 2019; Xue et al., 2022).

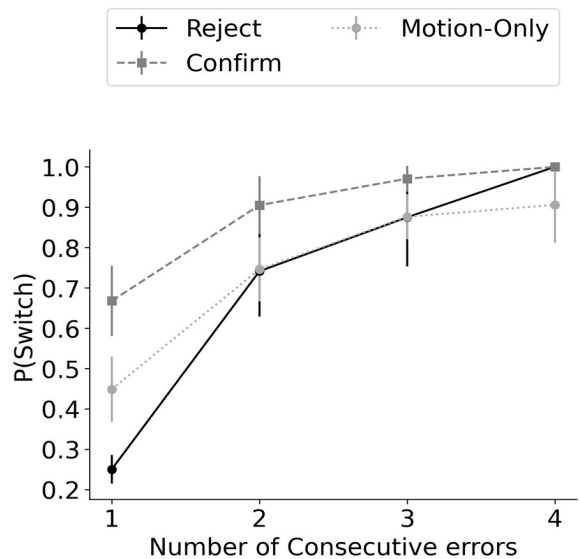

**Figure 3. Sequential accumulation of evidence for strategy switch.** Probability of switching grows with the number of consecutive errors. Compared to the Motion-Only condition, confirmed social advice increased the probability of switching.

Our experimental findings can help better understand the underlying computational mechanism in credit assignment in hierarchical decisions. Sequential sampling and accumulation of evidence has, for many decades, dominated the literature of perceptual decision making. (Purcell & Kiani, 2016) extended this framework to hierarchical decisions and proposed that higher order decisions are undertaken through a sequential sampling process (Heath, 1984; Wald, 1945). In order to decide if a higher level strategy switch is due, they proposed, participants treat the negative feedback as *evidence* for a change of environment. Importantly, this evidence is weighed by the participant's trial-by-trial expected accuracy. Following the general principle of sequential sampling, participants accumulate the switch evidence across trials until reaching a bound when they change environment. Consistent with this account, we found that the probability of switching increased with the number of consecutive errors (Figure 3). Moreover, and specific to our experiment here, we found that the evidence accumulation process across trials was modulated by social advice: Figure 3 demonstrates that compared to the Motion-Only condition, confirmed social advice increased the rate of accumulation of evidence across trials. Rejected advice had a more restricted impact on accumulation of evidence, primarily confined to reducing the impact of the first error. These findings are consistent with Figure 2B, where we saw that perceptual decisions that confirm (reject) the social advice are associated with higher (lower) confidence.

Our empirical findings demonstrated conclusively that social advice, independent of sensory input, influenced participants' confidence and thereby shaped their higher level strategy adjustments after negative feedback. The question whether confidence is the *exclusive* gatekeeper mediating this credit assignment remains open. This question connects to a controversial current topic in higher order cognition (Morales et al., 2018). At one extreme, it is possible that a *domain general* metacognitive mechanism integrates all evidence e.g., sensory, motor, economic value, social, moral, etc. into a singular confidence variable that is then utilized in higher order reasoning (Cole et al., 2013; De Gardelle & Mamassian, 2014; Fleming & Lau, 2014; Holroyd et al., 2005). On the other extreme, it is possible that each of these various, diverse streams of information gives rise to its own *domain specific* metacognition and corresponding confidence signal (Baird et al., 2013; Donoso et al., 2014; Fairhurst et al., 2018; McCurdy et al., 2013). Of course, these two extremes represent the theoretical boundaries and some in-between state of affairs is more likely in which the level and extent of domain generality/specificity could itself be a flexible source of neural and cognitive diversity between individuals and experimental contexts. An important avenue for the next step of the research from here would be to formulate these different alternatives as detailed computational models of the underlying mechanisms that could be compared against our open-access empirical data. For example, such models could investigate whether social advice impacts switching behavior directly or indirectly mediated by changes in motion confidence. What makes this question more interesting is that a number of recent studies have demonstrated that neural and computational mechanisms underlying confidence and metacognition are themselves subject to extensive diversity among individuals (Ais et al., 2016; Lehmann et al., 2022; Navajas et al., 2017).

Our study bridged sensory psychophysics and social psychology by investigating the effects of social context on perceptual performance. In sensory psychophysics, an observer is typically tested over thousands of trials, treating each dataset as a distinct experiment. In contrast, social psychology experiments involve fewer trials but more participants. Our challenge was balancing these approaches to examine the impact of social context on psychophysical performance. Due to the rarity of perceptual errors in high-coherence conditions, we concentrated on a few participants, each undergoing 2,400 trials, to ensure a sufficient number of rare error data. Individual results were consistent with the overall findings, despite some expected variations. Future studies could explore confidence in decision-making using alternative paradigms more aligned with social psychology standards to deepen understanding.

Needless to say, we admit that our experimental design and framework entail important reductions that do not fully capture the complexities of the real-world hierarchical decision-making such as buying a car where we started this paper. However, we would like to persuade the readers that our experimental setup does isolate a critical variable i.e., the role of confidence in credit assignment, in this mental process.

Additional investigations are needed to examine the generalizability of our findings across different contexts.

## Conclusion

We provided empirical evidence demonstrating that in a multilevel decision task, subjective confidence in low level perceptual decisions plays a key role in assignment of blame and credit to the appropriate level of decision hierarchy. Following negative feedback, strategic decisions were modulated by the impact of extra-sensory social advice on confidence.

# Supplementary Materials

**Decision confidence bridges credit assignment to levels of decision hierarchy**

Amir.M Mousavi Harris, Jamal Esmaily, Sajjad Zabbah, Reza Ebrahimpour, Bahador Bahrami

For code and data contact amir.harris1994@gmail.com

## Extended Materials and Methods

**Participants.** Following previous works (Purcell and Kiani, 2016; Sarrafyazd and Jayzayeri 2019), we treated each participant as a separate experimental test of our key hypothesis. Four participants, two men and two women (average±std: 24±5 years) with normal or corrected to normal vision were recruited. Participants were unaware of the aims of the experiment. After receiving the instructions, participants provided written consent. The Ethics Committee at the Iran University of Medical Science (Approval ID: IR.IUMS.REC.1400.1230) approved the study. Each participant undertook a total of 2400 trials in the main experiment and between 1000 to 1800 trials in the training phase.

**Motion Task.** Each trial began when the subject fixated on a small red fixation point (FP) with a diameter of 0.3° at the center of the screen. After a variable delay period (200–500 ms; truncated exponential distribution), two elongated bar targets (7° long, 0.75° wide) appeared on opposite sides of the screen, equidistant from the FP, at an eccentricity 8°. Subsequently, a dynamic random dot stimulus (for a detailed description of the task, refer to Shadlen and Newsome, 2001) was presented within a circular aperture of 5° centered on the FP. The duration of the motion stimulus on each trial was 500 ms.

The dots comprising the stimulus were white 4 × 4 pixels squares (0.096° × 0.096°) displayed on a black background. The dot density was set at 16.7 dots per square degree per second. The stimulus consisted of three independent sets of moving dots, shown in consecutive frames. Each set of dots was displayed for one video frame and then repositioned three video frames later (Δt = 40 ms). During repositioning, a subset of dots moved coherently to create apparent motion at a speed of 5° per second, while the remaining dots were randomly placed within the circular aperture.

Following the offset of the motion stimulus, a delay period (400–1,000 ms; truncated exponential distribution) was imposed before the Go signal, which was indicated by the FP disappearing. Subjects were instructed to gaze on the FP throughout the trial until the Go signal was presented. After the Go signal, subjects used a mouse to indicate their perceived direction of motion by moving the cursor to the corresponding choice target and clicking on it. Positive feedback was provided for correct responses, while negative feedback was given for incorrect responses, using distinct auditory cues. Each trial started when subjects successfully maintained their gaze within 3° around the FP.

Each target was assigned a specific color from a spectrum to incorporate a confidence rating component. The color spectrum ranged from green at one end, representing maximal certainty, to red at the other, indicating minimal certainty. The spectrum was divided into six discrete steps to represent varying confidence levels.

To manipulate the difficulty of the motion direction discrimination, we introduced random variations in the motion strength across trials. The motion strength was defined by the percentage of coherently displaced dots and varied between 0%, 3.2%, 6.4%, 12.8%, 25.6%, and 51.2%. For trials with 0% coherence, positive feedback was randomly given for half of the trials, while negative feedback was provided for the other half.

During the experiment, participants were positioned in an adjustable chair in a partially darkened room. Their chin and forehead were comfortably supported. They were in front of a Cathode Ray Tube (CRT) display monitor, specifically an EIZO FlexScan T966 model with a screen size of 20 inches. The monitor had a refresh rate of 75 Hz and a screen resolution of 1,600 × 1,200 pixels. The viewing distance between the participants and the monitor was 53 cm. The stimulus presentation was controlled using the Psychophysics Toolbox and Matlab software. Eye movements were tracked using a high-speed infrared camera (Eyelink; SR-Research) at a sampling rate of 1 kHz.

**Bandit (Changing Environment) Task.** After fixating on the fixation point (FP) and a variable delay period (200–500 ms; truncated exponential distribution), Participants were presented with two sets of choice targets, resulting in four-bar targets. These sets were positioned above and below the fixation point (FP), situated at 7° from the fixation point (FP), and arranged in a diamond pattern around the FP. Two pairs of targets were used: the upper left and lower right targets were oriented at 45°, while the upper right and lower left targets were oriented at 135°. Each set of targets corresponded to a specific motion direction: the right bar targets represented one direction, while the left targets represented the opposite direction. To differentiate between the two sets of choice targets, they are referred to as two distinct environments in this study. In each trial, only one of the two environments was correct. Participants were instructed to select the target corresponding to the correct motion direction and environment. We refer to the selection between left and right targets as the "direction choice" and between upper and lower targets as the "environment choice." The stability of an environment persisted for a certain number of consecutive trials, determined by a truncated geometric distribution with a range of 2 to 15 trials and a mean of 6. Subsequently, the environment changed. Participants were not explicitly informed about the correct environment or the timing of its change; instead, they had to discover it through trial and error. Positive feedback was provided only when the environment and direction choices were accurate. However, negative feedback was ambiguous, indicating an incorrect choice in the environment or direction. The auditory tones used for positive and negative feedback remained consistent with those employed during the direction discrimination training phase. Participants were explicitly instructed that their confidence ratings should reflect their confidence in their chosen motion direction and not their confidence in the chosen environment. In other words, the confidence ratings were meant to reflect the level of certainty regarding the direction choice, while the environment choice was not considered in the confidence assessment. Throughout the training, participants were informed that their objective was to maximize the proportion of correct trials. To achieve this, they were encouraged to identify environment changes as accurately and promptly as possible.

**Social Context.** participants were informed that they were paired with an anonymous partner for social context. However, they were paired with a computer-generated participant (CGP) designed to mimic their behavior from the changing environment task. The participants were not aware of this. The overall setup and stimulus presentation remained unchanged from the changing environment task. However, an additional delay of 700 ms was introduced after the presentation of the initial stimulus. Following this delay, a social cue was presented for 1,000 ms. This social cue indicates their partner's direction choice and confidence. In this panel, two horizontal bar targets were positioned at an eccentricity of 7° from the fixation point (FP), indicating the options for a leftward or rightward decision. The bar targets were accompanied by a white arrow, which served as a visual indicator of the chosen direction and the confidence level associated with the selected direction. After the presentation of the social cue, the response panel would be displayed, allowing participants to provide their responses based on the chosen direction and confidence level.

**Catch trials.** Following the provision of feedback in the social context, participants encountered a delay panel in approximately 10 percent of the trials. This delay panel lasted between 200 to 500 milliseconds. Subsequently, a new panel appeared, prompting participants to indicate whether their response to the Random Dot Motion (RDM) task aligned with the response provided by their partner. If the responses matched, participants were instructed to choose the green option. Conversely, if the responses did not match, they were asked to select the red option. This was used to ensure that subjects paid attention to the social information, and all subjects maintained an accuracy above 90%, indicating that they were consistently attending to the social information.

**Advice generation algorithm.** In our study, we were inspired by the methodology used by Bang et al. (2017) and Esmaily et al. (2023). We created four simulated partners for each participant based on their behavioral data from the Motion-Only session. These simulated partners were manipulated to have either high or low mean accuracy and high or low mean confidence. Each participant was paired with all four partner types—High Accuracy High Confidence (HAHC), High Accuracy Low Confidence (HALC), Low Accuracy High Confidence (LAHC), and Low Accuracy Low Confidence (LALC)—during a session, with each block featuring one partner (i.e., 300 trials per session for each partner type: HA, LA, HC, and LC). The order of partner pairings was randomized to prevent participants from identifying which partner they were paired with at any given time.

Initially, during the Motion-Only session, we calculated the participant's sensory noise (σ). We established specific thresholds to determine how confident they were in their responses. You can find more detailed information about this in the supplementary section of the sequential sampling model. For a more comprehensive understanding of this process, please refer to the research of Bang et al. (insert complete reference). The simulated partners were created to have either high accuracy (0.1× σ) or low accuracy (1.3×σ). The average confidence values were also determined based on the participant's data. For simulated partners with low confidence, the average confidence level was calculated using the participant's average confidence in trials with low coherence levels (3.2% and 6.4%). On the other hand, the mean confidence for simulated partners with high confidence was established using the participant's average confidence in trials with high coherence levels (25.6% and 51.2%).

We established the parameters for the simulated partners and then created trial-by-trial responses for each partner using the method outlined by Bang et al. (2017). The process of generating trial-by-trial responses involved several steps. First, we produced a sequence of coherence levels for each trial, with specific directions denoted as "+" for rightward and "-" for leftward. Next, we created a sequence of random values representing sensory evidence by drawing from a Gaussian distribution with a mean corresponding to the coherence levels and a variance of σ, representing the sensory noise. We used a set of thresholds from the participant's data in the Motion-Only condition to convert these random values into trial-by-trial responses and generate a partner with specific mean confidence. These thresholds were used to determine the participant's responses based on the sensory evidence for each trial. Additionally, we introduced a 5% chance of simulating lapses of attention and response errors by randomly selecting a response from a uniform distribution ranging from 0 to 1 in some trials.

In this model, the level of sensory noise (σ) determines the participant's sensitivity to the motion stimulus, while a set of 11 thresholds determines the participant's response distribution. These thresholds indicate both the decision made (via the sign of the threshold) and the confidence associated with that decision within the same distribution. The sensory evidence (x) on each trial is sampled from a Gaussian distribution, where x follows a normal distribution with a mean (s) and variance ($σ^2$). The mean (s) represents the motion coherence level and is selected uniformly from a set of values (S = {-0.512, -0.256, -0.128, -0.064, -0.032, 0.032, 0.064, 0.128, 0.256, 0.512}). The sign of s indicates the correct direction of motion, with positive values indicating rightward motion. The absolute value of s represents the strength of motion coherence. The standard deviation (σ) describes the level of sensory noise and is assumed to be constant across all stimuli. Our model considers the internal estimate of sensory evidence (z) equal to the raw sensory evidence (x). To determine the participant's sensitivity and response thresholds, we calculate the response distribution (r), which ranges from -6 to 6. The participant's confidence rating (c) is obtained by taking the absolute value of r, reflecting their confidence level in the decision. The actual decision is determined by the sign of r, with a positive sign indicating a rightward decision and a negative sign indicating a leftward decision.

The response distribution is mathematically described by Equation 1:

$$p_i = \begin{cases} p(z \leq \theta_{-6}) & i = -6 \\ p(\theta_{i-1} < z \leq \theta_i) & -6 < i \leq -1, \ 2 \leq i < 6 \\ p(\theta_{-1} < z \leq \theta_1) & i = 1 \\ p(z > \theta_5) & i = 6 \end{cases} \quad (1)$$

Using the values of θ and σ, we established a mapping between the internal estimate of sensory evidence (z) and the participants' responses (r). Specifically, we determined thresholds θi for each value of s in the set S, where i takes the values -6, -5, -4, -3, -2, -1, 1, 2, 3, 4, and 5. These thresholds were determined as follows:

$$\sum_{j \leq i} p_j = \frac{1}{10} \sum_{s \in S} \Phi\left(\frac{\theta_i - s}{\sigma}\right) \quad (2)$$

The S3 employs the Gaussian cumulative density function (Φ) to calculate the predicted response distribution, p(r = i|s), for each stimulus s ∈ S.

$$p(r = i|s) = \begin{cases} \Phi\left(\frac{\theta_{-6} - s}{\sigma}\right) & i = -6 \\ \Phi\left(\frac{\theta_{-i} - s}{\sigma}\right) - \Phi\left(\frac{\theta_{i-1} - s}{\sigma}\right) & -6 < i < 6 \\ 1 - \Phi\left(\frac{\theta_5 - s}{\sigma}\right) & i = 6 \end{cases} \quad (3)$$

Following this point, the model's accuracy can be calculated using Eq. 4.

$$a_{Subject} = \frac{\sum_{s \in S, s > 0} \sum_{i=1}^{6} p_{i,s} + \sum_{s \in S, s < 0} \sum_{i=-6}^{-1} p_{i,s}}{10} \quad (4)$$

Using the participant's accuracy, we can determine a set of optimal values for θ (response thresholds) and σ (sensory noise) to match the observed performance.

After determining the values of θ and σ, we can generate a confidence landscape with a predefined mean. It is important to note that an infinite number of confidence distributions can achieve the desired mean. Our focus is finding the maximum entropy distribution that satisfies two specific constraints: the mean confidence level being specified and the distribution summing up to 1. To accomplish this, we utilize the Lagrange multiplier (λ) method:

$$p_i = \frac{e^{i\lambda}}{\sum_{j=1}^{6} e^{i\lambda}} \quad (5)$$

The constraint is solved to determine the value of λ, which is then chosen to calculate the response distribution.

$$c = \frac{\sum_{j=1}^{6} j e^{j\lambda}}{\sum_{j=1}^{6} e^{j\lambda}} \tag{6}$$

By assuming symmetry around 0, we transformed the confidence distributions from 1 to 6 into response distributions from -6 to -1 and 1 to 6.

**Experimental Procedures**

**Eye Tracking.**

We recorded the eye movements using an EyeLink 1000 (SR-Research) device, which operated at a sampling rate of 1000 Hz and was managed by a separate, dedicated host PC. The device was configured in desktop mode and pupil-corneal reflection mode, focusing on data collection from the left eye. At the start of each block, participants underwent recalibration and validation using a 9-point schema displayed on the screen. Once the detection error was confirmed to be less than 0.5°, participants proceeded to the main task. Eye tracking was employed to ensure that participants maintained fixation on the designated fixation point during stimulus presentation. They were required to fixate on this point to initiate the trial.

**Training.** Initially, the subjects underwent training to become proficient in a motion discrimination task. The primary motion discrimination task training continued until subjects achieved high performance, maintaining a stable psychophysical threshold of 75% correct at 15% (or lower) coherence (Purcell & Kiani, 2016). Each trial started with the subjects looking at a centrally located fixation point (FP). After a short delay, two bar targets appeared on opposite sides of the screen, along with a random dot motion stimulus. The main goal for the subjects was to determine the overall direction of the motion (either left or right) while also assessing their confidence level in their decision (green indicated the highest confidence, and red indicated the lowest confidence). The presentation of the motion stimulus included different percentages of coherently moving dots (motion strength) from one trial to another, which adjusted the difficulty level of the motion direction discrimination task.

After a short delay, the FP was turned off. This signaled the participants to use their mouse to report the direction they thought the motion was going by moving the mouse pointer to the left or right. Different sounds were played to let them know if they made the right choice: a positive tone for a correct response and a negative tone for an incorrect response. Participants 1 to 4 took 1200, 1000, 1800, 1600 training trials, respectively.

**Motion-Only Condition.** After completing training in motion direction discrimination, participants were introduced to the main experiment, which consisted of two parts: the Motion-Only task and the Motion-Advice task. The Motion-Only task is identical to the changing environment task (Fig. 1.A) and is designed to engage participants in hierarchical decision-making. The experimental setup, motion stimulus, and sequence of events remained the same as during the training phase. However, a change was made to the number of choice bar targets presented to the subjects. Instead of a single pair of choice bar targets, subjects were presented with two pairs above and below the central fixation point (FP), resulting in four-bar targets. Each pair of targets corresponded to a distinct environment, with the right and left targets within each environment representing the two potential motion directions. Subjects received positive feedback when they correctly selected the target that matched the correct environment and motion direction. In other words, subjects provided simultaneous reports of their confidence in the direction choice and their selections for both the direction and environment choices. In this context, the selection between left and right targets was called the "direction choice," while the selection between upper and lower targets was called the "environment choice." The active environment remained fixed for several consecutive trials (ranging from 2 to 15 trials, with a mean of 6 trials following a truncated geometric distribution). Subsequently, it changed without an explicit cue (Fig. 1. A). To familiarize themselves with the task, participants completed 30 to 50 practice trials before the main data collection began. Each participant was asked to perform two sessions of the Motion Only task

**Motion-plus-Advice Condition.** After the Motion Only task, we added the social context to the changing environment task. The overall structure and stimulus presentation remained consistent with the changing environment task, but we introduced a 700-millisecond delay after the initial stimulus. A social cue then appeared for 1 second and participants were then asked to respond (Fig. 1. B). We included catch trials to ensure participants were attentive to the social cue. In 10% of the trials, an extra delay was added after participants responded, even though the cue was only shown for 1 second. Following this delay, the participants were asked to say whether their response to the Random Dot Motion (RDM) task matched the response given by their simulated partner (see Figure 1. C). Each participant was asked to perform two sessions of the Motion-Advice task.

# Extended Results

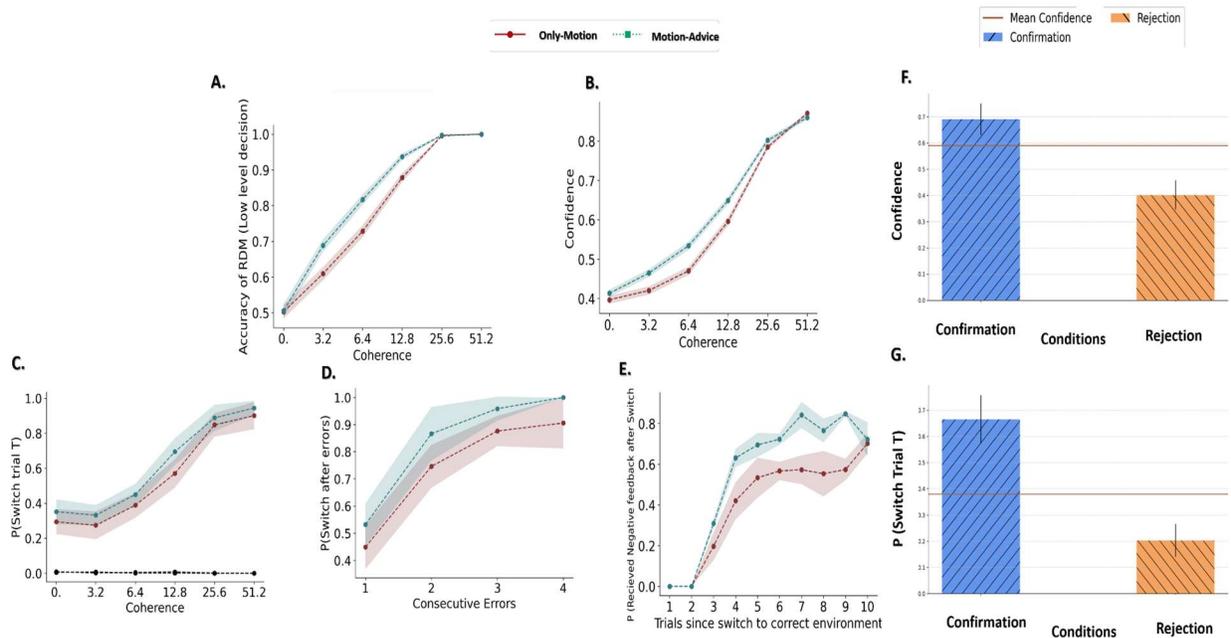

**Figure S1. Behavioral differences between one-cue and two-cue tasks**.(A) The probability of making the correct direction choice is higher in the two-cue task (green) compared to the one-cue task (red). (B) The reported confidence is also higher in the two-cue task (green) compared to the one-cue task (red). (C) The probability of switching after receiving negative feedback is higher in the two-cue task (green) compared to the one-cue task (red). The black lines and dots represent the probability of switching when receiving positive feedback. (D) Similar to (C), the probability of switching after consecutive errors is higher in the two-cue task (green) compared to the one-cue task (red). (E) The proportion of environment switches after negative feedback since the last correct switch is higher in the two-cue task (green) compared to the one-cue task (red). In all panels, red dots and lines represent one cue task, while blue indicates two. (F) This examines the effect of confirming or rejecting the partner's choice on the subject's confidence. It presents the mean confidence of subjects based on whether they confirm or reject the partner's choice. (G) This evaluates the impact of confirming or rejecting the partner's choice on the probability of switching. It shows the average probability of switching during confirmation and rejection trials. Blue bars indicate the mean probability during confirmation trials, while orange bars represent the mean during rejection trials. The green line represents the mean values from the one-cue task. Error bars represent SEM across subjects.

We evaluated the effects of motion strength and the subject's confidence on direction choices, independent of environment choices, using the following logistic regression (Eq.S7):

**P (Correct Direction)** = β0 + β1. Coherence (T) (<u>Motion-Only</u>) + β2. Confidence(T) + β3. Motion-Only or Advice-Motion (T) * Coherence (T)     (7)

Where P(Correct Direction) is the probability of choosing the correct motion direction on trial T. Coherence refers to motion strength, and Confidence is the subject's reported confidence. Motion-Only or Advice-Motion indicates whether the trial is Motion-Only or includes advice.

**Table S1 - Details of statistical results in RDM Accuracy.**

| Regressors | Estimate | SE | CI [0.025 0.975] | t-Stat | p-Value | No. Observations |
|---|---|---|---|---|---|---|
| **Const (β0)** | - 0.4810 | 0.073 | [-0.625 -0.337] | -6.556 | <0.001 | 9600 |
| **Coherence (β1)** | 17.7375 | 0.826 | [16.119 19.356] | 21.483 | <0.001 | 9600 |
| **Confidence (β2)** | 1.0186 | 0.118 | [0.787 1.250] | 8.623 | <0.001 | 9600 |
| **Interaction (Coherence) (β3)** | 0.2493 | 0.074 | [0.105 0.394] | 3.380 | 0.001 | 9600 |

We evaluated the effects of motion strength on the subject's confidence using the following multiple linear regression (Eq.S8):

***Confidence*** *= β0 + β1. Coherence (T)+ β2. Motion-Only or Advice-Motion (T) \* Coherence (T)*   (8)

Where Confidence is the subject's reported confidence.

**Table S2 - Details of statistical results in RDM Confidence.**

| Regressors | Estimate | SE | CI [0.025 0.975] | t-Stat | p-Value | No. Observations |
|---|---|---|---|---|---|---|
| **Const (β0)** | 0.4362 | 0.005 | [0.426 0.446] | 83.364 | <0.001 | 9600 |
| **Coherence (β1)** | 0.9320 | 0.018 | [0.897 0.967] | 52.844 | <0.001 | 9600 |
| **Interaction (coherence) (β2)** | 0.0305 | 0.006 | [0.018 0.043] | 4.950 | <0.001 | 9600 |

To measure how the previous trial influenced the decision to switch environment choices, we employed the following logistic regression (Eq.S9):

**P (switch) in (T) when receiving negative feedback in (T-1)** = $\beta_0$ + $\beta_1$. Coherence (T-1) + $\beta_2$. Confidence (T-1) + $\beta_3$. Motion-Only or Advice-Motion * Confidence (T-1) + $\beta_4$. Motion-Only or Advice-Motion * Coherence (T-1)  (9) where P (switch) is the probability that the environment choice on trial T does not match the environment choice on trial T - 1 (i.e., the subject switched environment choices) given positive (F+) or negative (F-) feedback on trial T - 1. Coherence and Confidence indicate the motion strength and subject's reported confidence on trial T - 1. Motion-Only or Advice-Motion indicates whether the trial is Motion-Only or includes advice.

**Table S3 - Details of statistical results in P(SW) with receiving one negative feedback.**

| Regressors | Estimate | SE | CI [0.025 0.975] | t-Stat | p-Value | No. Observations |
|---|---|---|---|---|---|---|
| Const ($\beta_0$) | - 2.0385 | 0.139 | [-2.310 -1.767] | -15.925 | <0.001 | 1331 |
| Coherence ($\beta_1$) | 4.6020 | 0.783 | [3.067 6.137] | 5.876 | <0.001 | 1331 |
| Confidence ($\beta_2$) | 2.0203 | 0.264 | [1.504 2.537] | 7.667 | <0.001 | 1331 |
| Interaction (coherence) ($\beta_3$) | 0.0401 | 0.014 | [0.014 0.067] | 2.964 | 0.003 | 1331 |
| Interaction (Confidence) ($\beta_4$) | 0.5783 | 0.268 | [0.054 1.103] | 2.160 | 0.031 | 1331 |

To assess the impact of the number preceding error trials on the choice to switch environments, we utilized the following logistic regression (Eq.S10):

**P (switch) in (T) when consecutive errors** = $\beta_0 + \beta_1 \cdot$ *Coherence (T-1)* $+ \beta_2 \cdot$ *Confidence(T-1)* $+ \beta_3 \cdot$ *Number of consecutive errors* $+ \beta_4 \cdot$ *Motion-Only or Advice-Motion * Number of consecutive errors* (10)

Where Number of consecutive errors is the count of consecutive negative feedback received prior to trial T.

**Table S4 - Details of statistical results in P(SW) with Consecutive errors**

| Regressors | Estimate | SE | CI [0.025 0.975] | t-Stat | p-Value | No. Observations |
|---|---|---|---|---|---|---|
| Const ($\beta_0$) | -2.7372 | 0.172 | [-3.074 -2.400] | -15.925 | <0.001 | 2032 |
| Coherence ($\beta_1$) | 6.2629 | 0.615 | [5.057 7.469] | 10.181 | <0.001 | 2032 |
| Confidence ($\beta_2$) | 1.9083 | 0.180 | [1.555 2.261] | 10.599 | <0.001 | 2032 |
| Consecutive Errors ($\beta_3$) | 1.0232 | 0.099 | [0.830 1.216] | 10.387 | <0.001 | 2032 |
| Interaction ($\beta_4$) | 0.3776 | 0.081 | [0.218 0.537] | 4.642 | <0.001 | 2032 |

To determine whether the subjects' decisions to switch environment choices were affected by the number of trials spent in the current environment, we employed the following regression (Eq.S11):

**P (switch) in (T) in urgency signal** = $\beta_0 + \beta_1 \cdot$ *Coherence (T-1)* $+ \beta_2 \cdot$ *Confidence (T-1)* $+ \beta_3 \cdot$ *Number of Stayed trials* $+ \beta_4 \cdot$ *Motion-Only or Advice-Motion * Number of Stayed trials* (11)

Where Number of Stayed trials represents the number of trials since the subject last switched to the new environment. In the task design, the environment duration was sampled from a truncated geometric distribution, which had a relatively flat hazard rate between trials 3 and 13, with the hazard rate increasing as it approached the truncation point. Subjects could adjust their switching behavior by learning the environment duration statistics, either through the hazard rate or the prior odds that the environment had changed since the last switch, based on their experience.

**Table S5 - Details of statistical results in P(SW) with Urgency signal**

| Regressors | Estimate | SE | CI [0.025 0.975] | t-Stat | p-Value | No. Observations |
|---|---|---|---|---|---|---|
| Const ($\beta 0$) | -3.2234 | 0.222 | [-3.659 -2.787] | -14.491 | <0.001 | 1337 |
| Coherence ($\beta 1$) | 5.4184 | 0.667 | [4.110 6.727] | 8.118 | <0.001 | 1337 |
| Confidence ($\beta 2$) | 2.0250 | 0.238 | [1.559 2.491] | 8.520 | <0.001 | 1337 |
| Urgency Signal ($\beta 3$) | 0.3077 | 0.035 | [0.240 0.375] | 8.916 | <0.001 | 1337 |
| Interaction ($\beta 4$) | 0.0825 | 0.025 | [0.033 0.132] | 3.292 | <0.001 | 1337 |

To evaluate the impact of the provided Advice on the subjects' confidence, we employed the following multiple linear regression (Eq.S12):

***Multiple Linear Regression on Confidence in (T)** = β0 + β1. Coherence (T) (<u>Motion-Only</u>) + β2. Confidence of Agent (T) (<u>Advice Confidence</u>) + β3. Accuracy Type of Agents + β4. Confidence Type of Agents + β5. Confirmation Vs Rejection (T) (<u>Advice-Motion</u>)*                    (12)

Here, we evaluate the partner's confidence, the type of agent (high or low accuracy/confidence), and whether they confirmed or rejected the partner's direction choice on subject's confidence.

**Table S6 - Details of statistical results of Confidence in Confirmation or Rejection**

| Regressors | Estimate | SE | CI [0.025 0.975] | t-Stat | p-Value | No. Observations |
|---|---|---|---|---|---|---|
| Const (β0) | 0.2785 | 0.012 | [0.255 0.303] | 22.753 | <0.001 | 3150 |
| Coherence (β1) | 0.5945 | 0.029 | [0.537 0.652] | 20.208 | <0.001 | 3150 |
| Confidence (Partner) (β2) | 0.0369 | 0.003 | [0.031 0.043] | 11.603 | <0.001 | 3150 |
| Partner Type (Accuracy) (β3) | 0.0067 | 0.008 | [-0.009 0.022] | 0.826 | 0.409 | 3150 |
| Partner Type (Confidence) (β4) | -0.0112 | 0.010 | [-0.030 0.008] | -1.144 | 0.253 | 3150 |
| Confirmation Or Rejection (β5) | 0.1247 | 0.012 | [0.102 0.148] | 10.695 | <0.001 | 3150 |

We also assessed the impact of the provided advice on the probability of switching, using the following logistic regression (Eq.S13):

***Logistic Regression of P (switch) in (T) when receiving negative feedback in (T-1)*** $= β0 + β1.$ *Coherence (T-1) (<u>Motion-Only</u>) + β2. Confidence of Subject (T-1) + β3. Accuracy Type of Agent + β4. Confidence Type of Agent+ β5. Confidence of Agent (T-1) (<u>Advice Confidence</u>) + β6. Confirmation Vs Rejection (T-1) (<u>Advice-Motion</u>)* (13)

Where we evaluate the partner's confidence, the type of agent (high or low accuracy/confidence), the subject's confidence, and whether they confirmed or rejected the partner's direction choice, all in relation to the probability of switching.

**Table S7 - Details of statistical results of P(sw) in Confirmation or Rejection**

| Regressors | Estimate | SE | CI [0.025 0.975] | t-Stat | p-Value | No. Observations |
|---|---|---|---|---|---|---|
| **Const (β0)** | -2.8414 | 0.257 | [-3. 346 -2.337] | -11.044 | <0.001 | 890 |
| **Coherence (β1)** | 5.2948 | 1.073 | [3.191 7.398] | 4.933 | <0.001 | 890 |
| **Confidence (subject) (β2)** | 2.3417 | 0.333 | [1.688 2.995] | 7.024 | <0.001 | 890 |
| **Confidence (Partner) (β3)** | 0.0489 | 0.064 | [-0.076 0.174] | 0.765 | 0.444 | 890 |
| **Partner Type (Accuracy) (β4)** | 0.3696 | 0.167 | [0.043 0.696] | 2.217 | 0.027 | 890 |
| **Partner Type (Confidence) (β5)** | -0.1518 | 0.201 | [-0.545 0.241] | -0.757 | 0.449 | 890 |
| **Confirmation Or Rejection** | 1.3796 | 0.189 | [1.010 1.749] | 7.310 | <0.001 | 890 |

| | | | | | |
|---|---|---|---|---|---|
| (β6) | | | | | |

**Table S8 - Details of statistical results of Confidence in Rejection**

To test whether subjects utilize social confidence advice in their confidence for switching during rejection, we applied the following regression:

***Multiple Linear Regression on Confidence** = β0 + β1. Coherence (T) (<u>Motion-Only</u>) + β2. Confidence of Agent (T) (<u>Advice Confidence</u>)* (14)

| Regressors | Estimate | SE | CI [0.025 0.975] | t-Stat | p-Value | No. Observations |
|---|---|---|---|---|---|---|
| Const (β0) | 0.363 | 0.027 | [0.310 0.416] | 13.541 | <0.001 | 297 |
| Coherence (β1) | 0.338 | 0.309 | [-0.270 0.947] | 1.095 | 0.275 | 297 |
| Confidence (Partner) (β2) | 0.006 | 0.008 | [-0.010 0.023] | 0.767 | 0.443 | 297 |

**Table S9 - Details of statistical results of Confidence in Confirmation**

We applied the same equation (Eq. 14) for the confirmation conditions as well.

| Regressors | Estimate | SE | CI [0.025 0.975] | t-Stat | p-Value | No. Observations |
|---|---|---|---|---|---|---|
| Const (β0) | 0.337 | 0.022 | [0.294 0.382] | 15.138 | <0.001 | 593 |
| Coherence (β1) | 0.516 | 0.069 | [0.381 0.652] | 7.482 | <0.001 | 593 |
| Confidence (Partner) | 0.058 | 0.006 | [0.046 0.071] | 9.141 | <0.001 | 593 |

| (β2) | | | | | | |

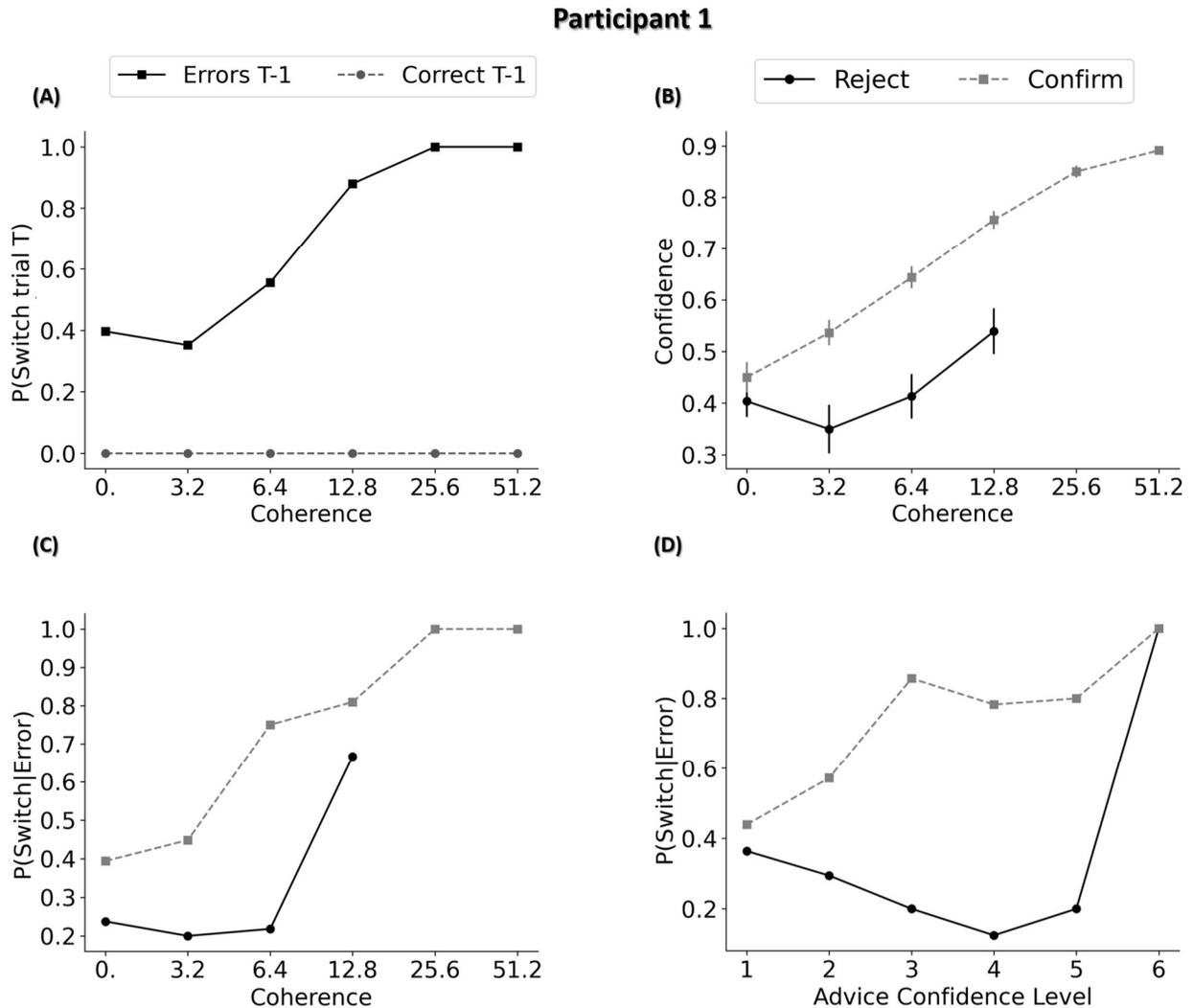

Figure S2. Behavioral Data for participant 1. (A) Probability of switching when receiving one negative feedback increases with the coherence level. It grows as a function of the coherence level. (B) The effect of confirming or rejecting the partner's choice influences the subjects' confidence. Subjects' confidence levels vary depending on whether they confirm or reject their partner's choice. (C) The impact of confirming or rejecting the partner's choice on the probability of switching reveals that the probability of switching increases during confirmation and decreases during rejection. (D) The effect of the partner's confidence in the Motion Advice task on switching probability shows that the likelihood of switching varies between confirmation and rejection.

**Participant 2**

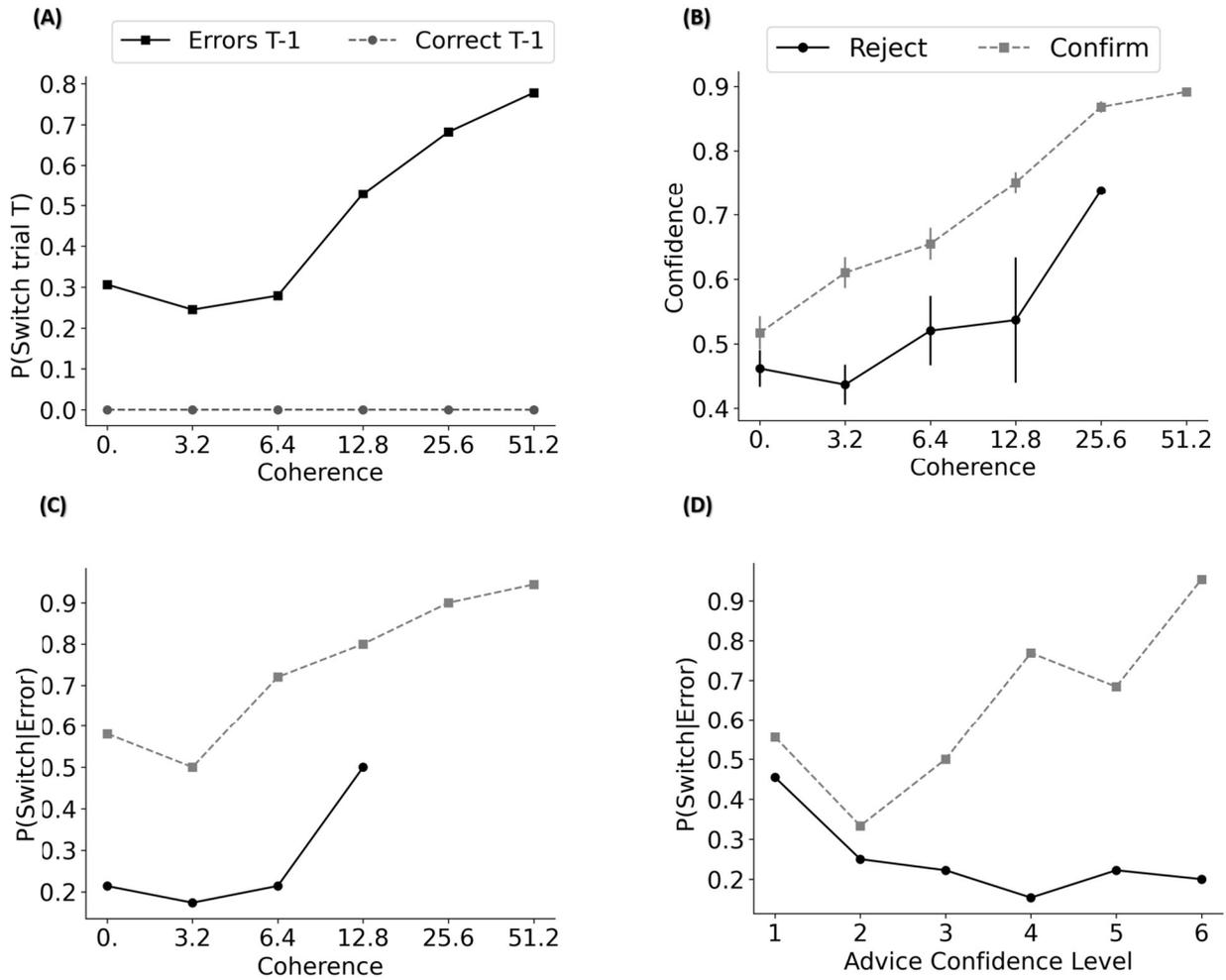

**Figure S3. Behavioral Data for participant 2.** Conventions are the same as Figure S2.

# Participant 3

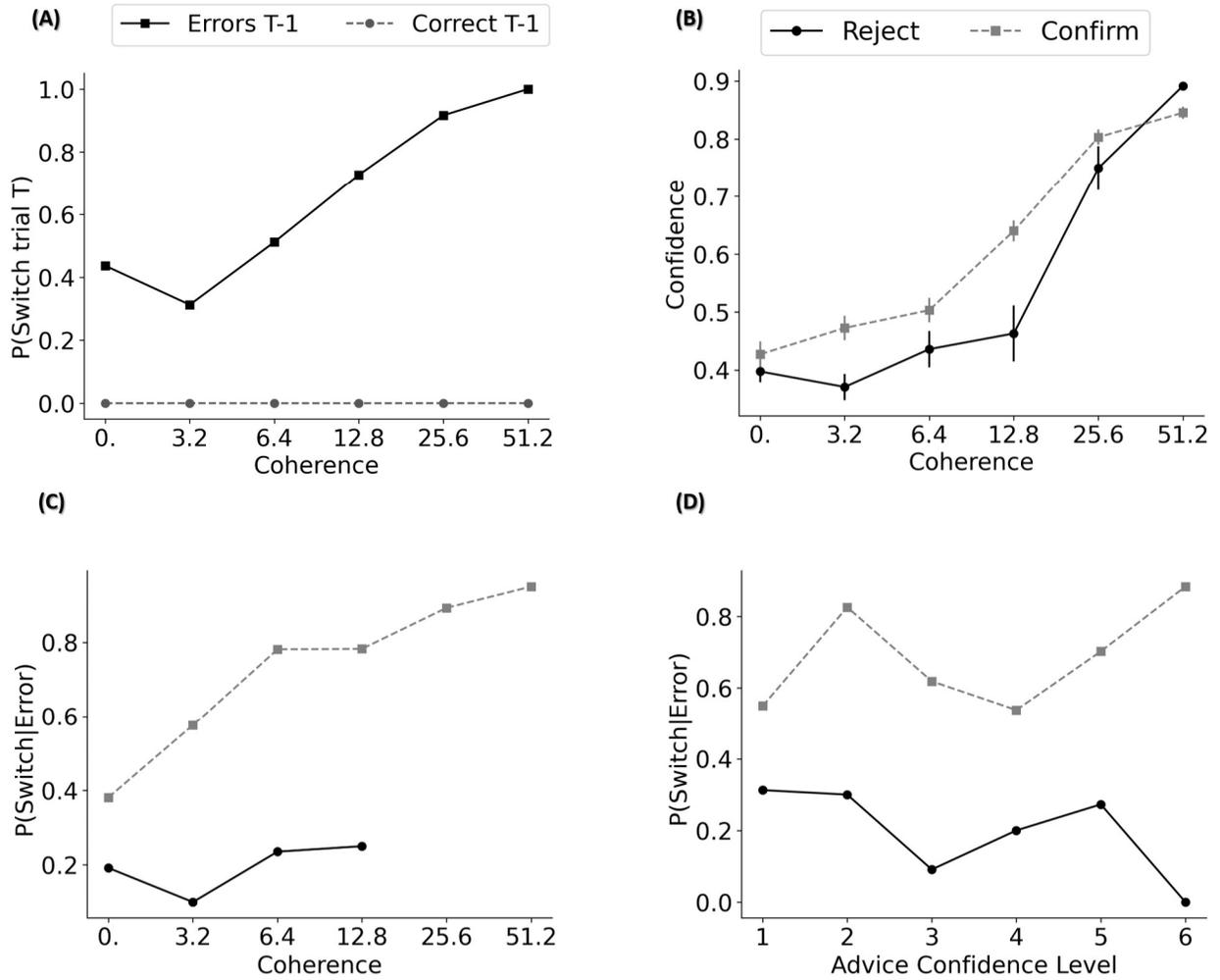

**Figure S4. Behavioral Data for participant 3.** Conventions are the same as Figure S2.

# Participant 4

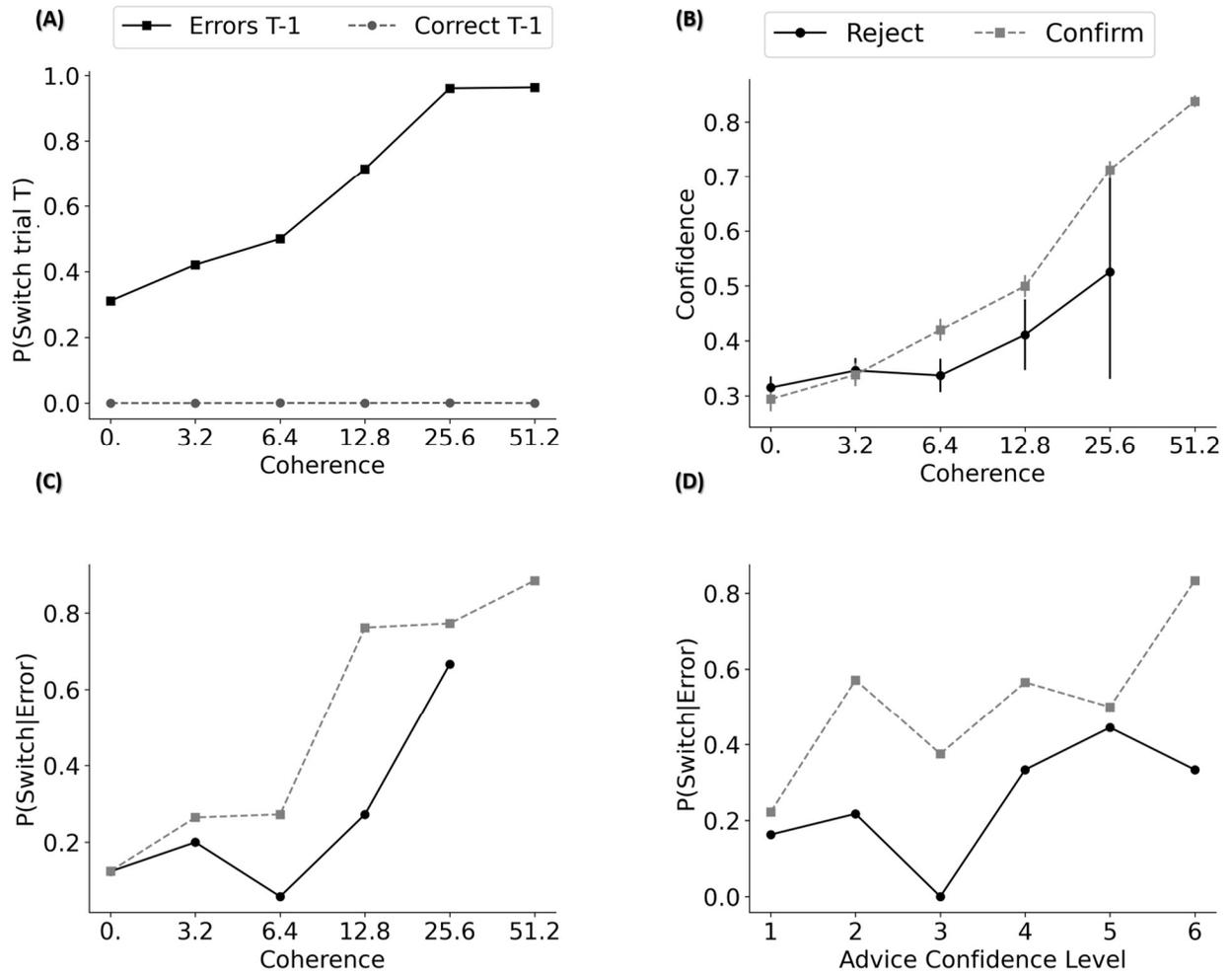

**Figure S5. Behavioral Data for participant 4.** Conventions are the same as Figure S2.

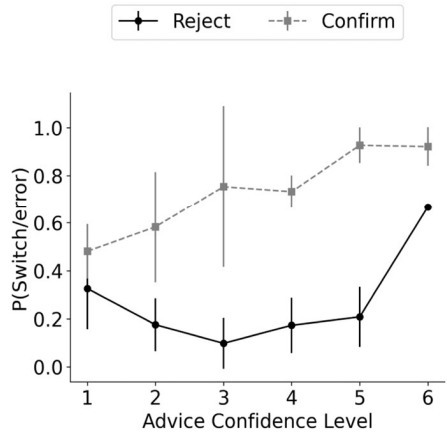 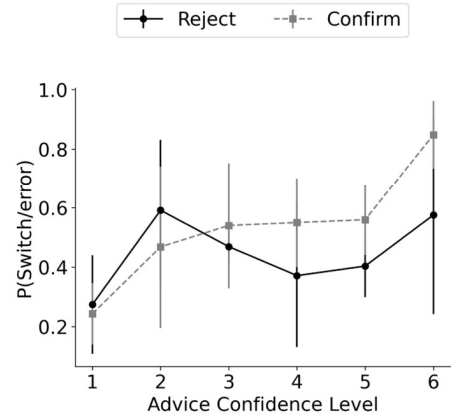

**Figure S6. Impact of advice accuracy on strategic decisions.** Left. High Accuracy partner. Switching probability increases with confirmed (but not rejected) advice confidence. Right. Low Accuracy partner. The switching probability for both confirmed and rejected cases is identical. Error bars are Standard Error of Mean.

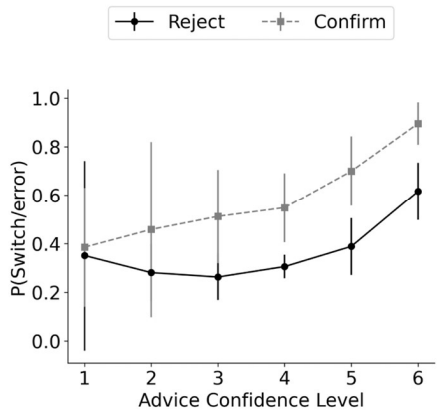 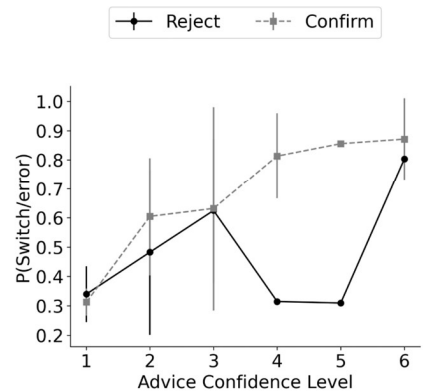

**Figure S7. Impact of advice confidence on strategic decisions.** Left. High Confidence partner. Switching probability increases with confirmed (but not rejected) advice confidence. Right. Low Confidence partner. The switching probability for both confirmed and rejected cases is almost identical. Error bars are Standard Error of Mean.

**Table S10 - Details of statistical results of participant 1.**

| Equations | Regressors | Estimate | SE | t-Stat | p-Value | No. Observations |
|---|---|---|---|---|---|---|
| Eq. S7 | Coherence | 17.866 | 1.791 | 9.976 | <0.001 | 2400 |
| | Confidence (subject) | 1.206 | 0.207 | 5.823 | <0.001 | 2400 |
| | Social effect on Accuracy | 0.069 | 0.026 | 2.661 | 0.008 | 2400 |
| Eq. S8 | Social effect on Confidence | 0.075 | 0.011 | 6.576 | <0.001 | 2400 |
| Eq. S9 | Social effect on P (sw) in T-1 | 1.426 | 0.264 | 5.398 | <0.001 | 493 |
| Eq. S10 | Social effect on Consecutive errors | 1.145 | 0.220 | 5.218 | <0.001 | 625 |
| Eq. S11 | Social effect on Urgency signal | 0.316 | 0.053 | 6.021 | <0.001 | 466 |
| Eq. S12 | Confirmation Vs Rejection | 0.175 | 0.021 | 8.405 | <0.001 | 2400 |
| | Partner's Confidence | 0.054 | 0.005 | 7.935 | <0.001 | 2400 |
| | Confirmation Vs Rejection | 0.830 | 0.365 | 2.273 | 0.023 | 248 |

| Eq. S13 | Partner's Confidence | -0.121 | 0.119 | -1.017 | 0.309 | 248 |
| Eq. S14 | Partner's Confidence (Reject) | 0.033 | 0.019 | 1.803 | 0.076 | 70 |
| | Partner's Confidence (Confirm) | 0.078 | 0.012 | 6.609 | <0.001 | 178 |

**Table S11 - Details of statistical results of participant 2.**

| Equations | Regressors | Estimate | SE | t-Stat | p-Value | No. Observations |
|---|---|---|---|---|---|---|
| Eq. S7 | Coherence | 17.082 | 1.655 | 10.325 | <0.001 | 2400 |
| | Confidence (subject) | 0.839 | 0.185 | 4.530 | <0.001 | 2400 |
| | Social effect on Accuracy | 0.066 | 0.024 | 2.792 | 0.005 | 2400 |
| Eq. S8 | Social effect on Confidence | 0.044 | 0.014 | 3.229 | 0.001 | 2400 |
| Eq. S9 | Social effect on P (sw) in T-1 | 0.367 | 0.216 | 1.703 | 0.089 | 449 |
| Eq. S10 | Social effect on Consecutive errors | 0.567 | 0.159 | 3.558 | <0.001 | 676 |
| Eq. S11 | Social effect on Urgency signal | 0.499 | 0.235 | 2.124 | 0.034 | 421 |
| Eq. S12 | Confirmation Vs Rejection | 0.158 | 0.026 | 6.101 | <0.001 | 2400 |
| | Partner's Confidence | 0.017 | 0.006 | 2.88 | 0.004 | 2400 |
| Eq. S13 | Confirmation Vs Rejection | 1.000 | 0.330 | 3.031 | 0.002 | 249 |
| | Partner's Confidence | 0.029 | 0.095 | 0.306 | 0.759 | 249 |
| Eq. S14 | Partner's Confidence (Reject) | -0.012 | 0.018 | -0.676 | 0.501 | 86 |
| | Partner's Confidence (Confirm) | 0.037 | 0.012 | 3.155 | 0.002 | 168 |

**Table S12 - Details of statistical results of participant 3.**

| Equations | Regressors | Estimate | SE | t-Stat | p-Value | No. Observations |
|---|---|---|---|---|---|---|

| Equations | Regressors | Estimate | SE | t-Stat | p-Value | No. Observations |
|---|---|---|---|---|---|---|
| Eq. S7 | Coherence | 16.497 | 1.190 | 13.864 | <0.001 | 2400 |
| | Confidence (subject) | 1.219 | 0.309 | 3.947 | <0.001 | 2400 |
| | Social effect on Accuracy | 0.489 | 0.115 | 4.270 | <0.001 | 2400 |
| Eq. S8 | Social effect on Confidence | 0.018 | 0.010 | 1.846 | 0.065 | 2400 |
| Eq. S9 | Social effect on P (sw) in T-1 | 0.359 | 0.200 | 1.793 | 0.073 | 528 |
| Eq. S10 | Social effect on Consecutive errors | 0.228 | 0.179 | 1.278 | 0.201 | 779 |
| Eq. S11 | Social effect on Urgency signal | 0.023 | 0.046 | 0.506 | 0.613 | 421 |
| Eq. S12 | Confirmation Vs Rejection | 0.058 | 0.018 | 3.227 | 0.001 | 2400 |
| | Partner's Confidence | 0.014 | 0.004 | 3.373 | 0.001 | 2400 |
| Eq. S13 | Confirmation Vs Rejection | 1.140 | 0.312 | 3.651 | <0.001 | 234 |
| | Partner's Confidence | 0.010 | 0.089 | 0.113 | 0.910 | 234 |
| Eq. S14 | Partner's Confidence (Reject) | 0.002 | 0.009 | 0.230 | 0.818 | 91 |
| | Partner's Confidence (Confirm) | 0.035 | 0.008 | 4.304 | <0.001 | 171 |

Table S13 - Details of statistical results of participant 4.

| Equations | Regressors | Estimate | SE | t-Stat | p-Value | No. Observations |
|---|---|---|---|---|---|---|
| Eq. S7 | Coherence | 13.572 | 1.228 | 11.053 | <0.001 | 2400 |
| | Confidence (subject) | 0.769 | 0.154 | 5.000 | <0.001 | 2400 |
| | Social effect on Accuracy | 0.055 | 0.017 | 3.187 | 0.001 | 2400 |
| Eq. S8 | Social effect on Confidence | 0.033 | 0.014 | 2.395 | 0.017 | 2400 |
| Eq. S9 | Social effect on P (sw) in T-1 | 0.353 | 0.242 | 1.463 | 0.143 | 536 |
| Eq. S10 | Social effect on Consecutive errors | 0.305 | 0.178 | 1.718 | 0.086 | 903 |
| Eq. S11 | Social effect on Urgency signal | 0.479 | 0.202 | 2.376 | 0.039 | 646 |
| Eq. S12 | Confirmation Vs Rejection | 0.048 | 0.024 | 2.069 | 0.039 | 2400 |
| | Partner's Confidence | 0.020 | 0.006 | 3.236 | 0.001 | 2400 |
| Eq.S13 | Confirmation Vs Rejection | 2.405 | 0.869 | 2.768 | 0.006 | 262 |
| | Partner's Confidence | -0.056 | 0.216 | -0.261 | 0.794 | 262 |
| Eq. S14 | Partner's Confidence (Reject) | 0.005 | 0.025 | 0.215 | 0.830 | 94 |
| | Partner's Confidence (Confirm) | 0.065 | 0.014 | 4.678 | <0.001 | 150 |